\documentclass[twocolumn,prl,showpacs,amsfonts,amsmath,assymb,eufrak,longbibliography]{revtex4-1}
\usepackage{epsfig}
\usepackage{color}
\usepackage{bm}
\usepackage[
colorlinks=true,
pdfstartview=FitV,
linkcolor=blue,
citecolor=magenta,
urlcolor=blue,
bookmarks=true,
bookmarksnumbered=true,
pdftitle={},
pdfauthor={}
]{hyperref}
\usepackage{umoline}
\usepackage{braket}
\usepackage{epsfig,amsopn}
\usepackage{graphicx}
\usepackage{amsmath,amssymb}

\usepackage{calc}

\newcommand{\sectionprl}[1]{{\em #1}\/.---}

\newtheorem{theorem}{Theorem}
\newtheorem{lemma}{Lemma}

\newtheorem{statement}[theorem]{Statement}
\usepackage{color}
\usepackage[usenames,svgnames]{xcolor}

\newcommand{\vac}{{\rm vac}}


\begin{document}

\preprint{APS/123-QED}

\newcommand{\titlename}{Fate of measurement-induced phase transition in long-range interactions}

\preprint{APS/123-QED}

\title{\titlename}

\author{Takaaki Minato$^{1}$, Koudai Sugimoto$^{1}$, Tomotaka Kuwahara$^{2}$, and Keiji Saito$^{1}$ }
\affiliation{$^{1}$Department of Physics, Keio University, Hiyoshi, Kohoku-ku, Yokohama 223-8522, Japan}%
\affiliation{$^{2}$ Mathematical Science Team, RIKEN Center for Advanced Intelligence Project (AIP), 1-4-1 Nihonbashi, Chuo-ku, Tokyo 103-0027, Japan}%

\date{\today}%

\begin{abstract}
We consider quantum many-body dynamics under quantum measurements, where the measurement-induced phase transitions (MIPs) occur when changing the frequency of the measurement. In this work, we consider the robustness of the MIP for long-range interaction that decays as $r^{-\alpha}$ with distance $r$. The effects of long-range
interactions are classified into two regimes: (i) the MIP is observed $(\alpha > \alpha_c)$, and (ii) the MIP is absent even for arbitrarily strong measurements $(\alpha<\alpha_c)$. Using fermion models, we demonstrate both regimes in integrable and non-integrable cases. We identify the underlying mechanism and propose sufficient conditions to observe the MIP, that is, $\alpha > d/2+1$ for general bilinear systems and $\alpha > d+1$ for general non-integrable systems ($d$: spatial dimension). Numerical calculation indicates that these conditions are optimal.

\end{abstract}

\maketitle


\sectionprl{Introduction}
Understanding the general properties and finding new phenomena regarding the time evolution of quantum entanglement in quantum many-body systems is a critical subject in physics. Recently, novel dynamic phase transitions in quantum entanglement have been discovered in the presence of quantum measurements ~\cite{Li_2018,Chan_2019,Skinner_2019,Li_2019,gullans2019purification,gullans2019scalable,jian2019measurementinduced,Bao_2020,Choi_2020,Szyniszewski_2019,fan2020selforganized,vijay2020measurementdriven,lavasani2020measurementinduced,sang2020measurement,ippoliti2020entanglement,Tang_Zhu_2020,goto2020measurementinduced,fuji2020measurement,alberton2021entanglement,fidkowski2020dynamical,Turkeshi_2020,lang2020entanglement}. In general, the bipartite entanglement entropy of isolated systems grows over time and eventually reaches the order of the system size. Conversely, projective quantum measurements suppress entanglement growth, such as in the quantum Zeno effect under continuous measurement~\cite{wiseman2009quantum}. As a result of this competition, with increasing measurement amplitude (frequency) in non-integrable systems, the bipartite entanglement entropy in the long-time limit shows a transition from the order of the system size (the volume law phase) to the order of the boundary area (the area law phase). This phenomenon is now referred to as measurement-induced phase transition (MIP).

MIPs have been intensively studied in various systems, such as quantum circuit models \cite{Li_2018,Chan_2019,Skinner_2019,Li_2019,gullans2019purification,gullans2019scalable,jian2019measurementinduced,Bao_2020,Choi_2020,Szyniszewski_2019,fan2020selforganized,vijay2020measurementdriven,lavasani2020measurementinduced,sang2020measurement,ippoliti2020entanglement}, cold atomic systems \cite{Tang_Zhu_2020,goto2020measurementinduced,fuji2020measurement,alberton2021entanglement},
and quantum spin systems \cite{Turkeshi_2020,lang2020entanglement}. We emphasize that the MIP generally occurs irrespective of integrability/non-integrability. It has been recently found that free-fermion systems also show the MIP, i.e., a transition between the phase of the entanglement entropy with the order of the logarithmic system size (the sub-volume law phase) and the area law phase \cite{alberton2021entanglement}. The MIP is believed to be a ubiquitous phenomenon in isolated many-body quantum systems.

\begin{figure}[t]
\centering
{
  \includegraphics[clip, scale=0.25]{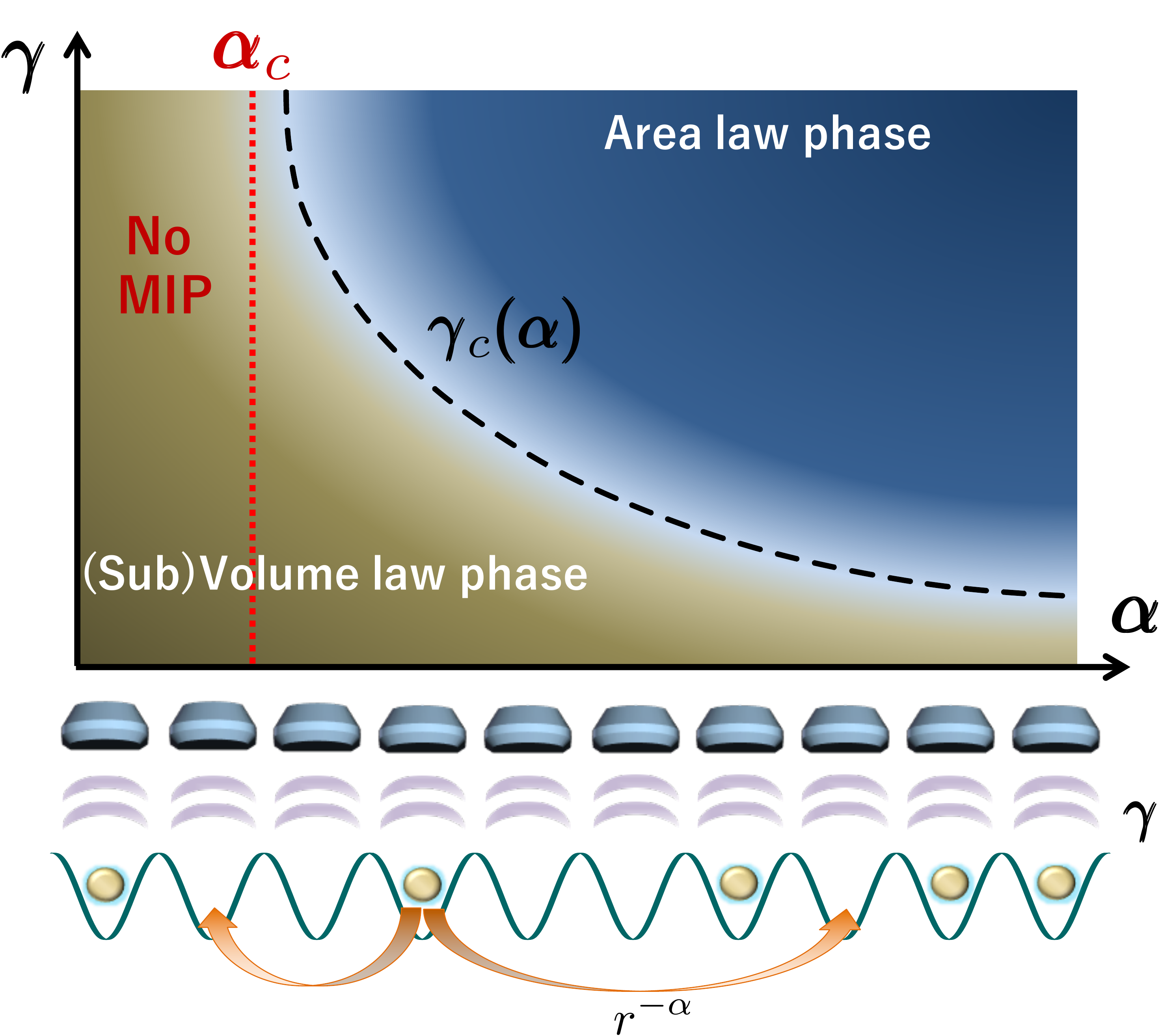}
\caption{Schematic of findings. We consider fermion systems with long-range interactions, where each site is constantly measured by the amplitude (frequency) $\gamma$.
  The function $\gamma_c(\alpha)$ is the critical measurement amplitude (see Fig.~{\ref{fig3}}~(b) for the free fermion case). In the regime $\alpha <\alpha_c$, the MIP does not exist.
  See Statement~\ref{cond_MIP} for sufficient conditions of $\alpha$ for the existence of the MIP.
    \label{fig1} }
  }
\end{figure}

In this paper, we consider the MIP in long-range interacting systems to understand the mechanism more deeply. Here, long-range interaction means that the amplitude of interaction decays as $r^{-\alpha}$, where $r$ is the distance between particles (see also~\footnote{As in Ref.~\cite{nahum2020measurement}, a quantum circuit model with randomly selected networks can also be called a long-range interacting model. However, it is not relevant to ours because the system we address here is a Hamiltonian system, where each particle simultaneously interacts with the rest of the particles depending on the distance.}). As conceptually established in statistical mechanics \cite{lifshitz2013statistical}, phase transitions generally depend on the interaction range, dimensionality, and types of interactions.
Various studies have shown that physical properties change qualitatively under long-range interactions. Examples include the static properties of the equilibrium phase~\cite{dyson1969existence,thouless1969long,kosterlitz1976phase}, ground state~\cite{kuwahara2020area,koffel2012entanglement,vodola2014kitaev,vodola2015long}, and dynamic properties~\cite{chen2019finite,tran2020hierarchy,kuwahara2020strictly,zhou2020operator,tamaki2020energy,PhysRevLett.126.030604,nandkishore2017many}. Thus, it is natural to ask whether long-range interactions influence the physics of MIP.

The long-range interaction immediately propagates the quantum information to particles with arbitrary distances, and hence, the entanglement growth should be enhanced. From this viewpoint, one anticipates nontrivial competition between quantum measurement and long-range interaction strength. We here address the following questions: (i) is there a possibility for the absence of the MIP ?; and (ii) what are the conditions for the existence of the MIP ?

The primary obstacle to addressing these questions lies in the fact that the dynamics under quantum measurements are highly nonlinear~\cite{wiseman2009quantum}, which makes the analyses difficult even numerically. We note that simple Clifford circuit models have been employed in many studies so far to overcome this difficulty\cite{Li_2018,Chan_2019,Li_2019,gullans2019purification,gullans2019scalable,vijay2020measurementdriven,lavasani2020measurementinduced,sang2020measurement,ippoliti2020entanglement}. Following this spirit, we use a simple toy model to grasp the essence. We start with a simple fermion model to address question (i). We then identify the physical mechanism to make a statement applicable to generic systems addressing question (ii), where sufficient conditions to observe the MIP in generic systems are proposed.
Then, we obtain several physical pictures, as summarized schematically in Fig.~\ref{fig1}.

\sectionprl{Model}
To obtain the essential physics of the effect of long-range interaction in the MIP, we consider the following simple long-range Hamiltonian:
\begin{align}
  H &= \sum_{j=1}^L \sum_{r=1}^{L/2}{1\over r^{\alpha}} \left[ -c_{j+r}^{\dagger}c_j
      -c_{j}^{\dagger}c_{j+r} + V n_{j+r} n_j  \right] \, , \label{hamil}
\end{align}
where $c_{j}$ and $c_{j}^{\dagger}$ are the annihilation and creation operators of the spinless fermion at site $j$, and $n_j =c_j^{\dagger} c_j$, respectively. The parameter $\alpha$ is the degree of long-range interaction. We impose the periodic boundary condition for the total system size $L$, that is, $c_{j+L}=c_{j}$. We set the Neel state to the initial state, that is, $\psi (t=0)=\prod_{i=1}^{L/2} c^{\dagger}_{2i-1} |\vac \rangle$, where $|\vac\rangle$ is the vacuum state. Note that the total number of fermions is fixed at $L/2$ at all times. We perform a quantum measurement uniformly for all sites with a finite measurement amplitude (frequency) $\gamma$. Then, the time evolution of the wave function is described by the standard quantum jump process \cite{wiseman2009quantum,fuji2020measurement}, that is,
\begin{align}
  d | \psi (t) \rangle &= -i H | \psi (t) \rangle dt \nonumber \\ &+ \sum_{j=1}^L
                         \left[
                         {c_j^{\dagger} c_j | \psi (t) \rangle
           \over \sqrt{\langle \psi (t) | n_j | \psi (t) }} - | \psi (t) \rangle
                         \right] d w_j (t) \, ,
\end{align}
where $d w_j (t)$ takes $0$ or $1$ obeying
the site-independent Poisson process, that is, $\langle\langle d w_j (t) \rangle\rangle = \gamma dt$. Here, $\langle\langle ... \rangle\rangle$ is the noise average. As a result of the measurement process, there are many trajectories of the wave functions starting from the fixed initial state. Hence, we need to take an average over many trajectories to examine the statistical properties of any observables. The main physical quantity we address is entanglement entropy. Let us divide the system into two subsystems $A$ and $B$, which have sizes $\ell$ and $L-\ell$ $(\ell \le L/2)$, respectively. Then, the entanglement entropy is defined as:
\begin{align}
  S_{\ell} &:= -{\rm Tr} ( \rho_{A} \log \rho_{A}).
\end{align}
where $\rho_A$ is the reduced density matrix of subsystem $A$ for a given wave function $|\psi \rangle$, that is, $\rho_A ={\rm Tr}_B (|\psi \rangle \langle \psi |)$, where ${\rm Tr}_B$ is the partial trace with respect to part $B$. We discuss the trajectory average $\bar{S}_{\ell}$, employing a sufficient number of trajectories.
We also compute the mutual information as another indicator to detect the MIP. To this end, we divide the total system into four subsystems in the order of “a”, “b”, “c” and “d” along the system (hence, the regions “a” and “d” contact with each other on the ring geometry).
We then consider the mutual information between the region “a” and “c”, $I(\gamma , \alpha) = S_a + S_c - S_{ac}$, where $S_a$, $S_c$ and $S_{ac}$ are the entanglement entropy for the regions “a”, “c”, and “a”+“c”, respectively. We denote the average values of mutual information by $\bar{I}(\gamma , \alpha )$.

\begin{figure}[t]
\centering
{
  \includegraphics[clip, width=\linewidth]{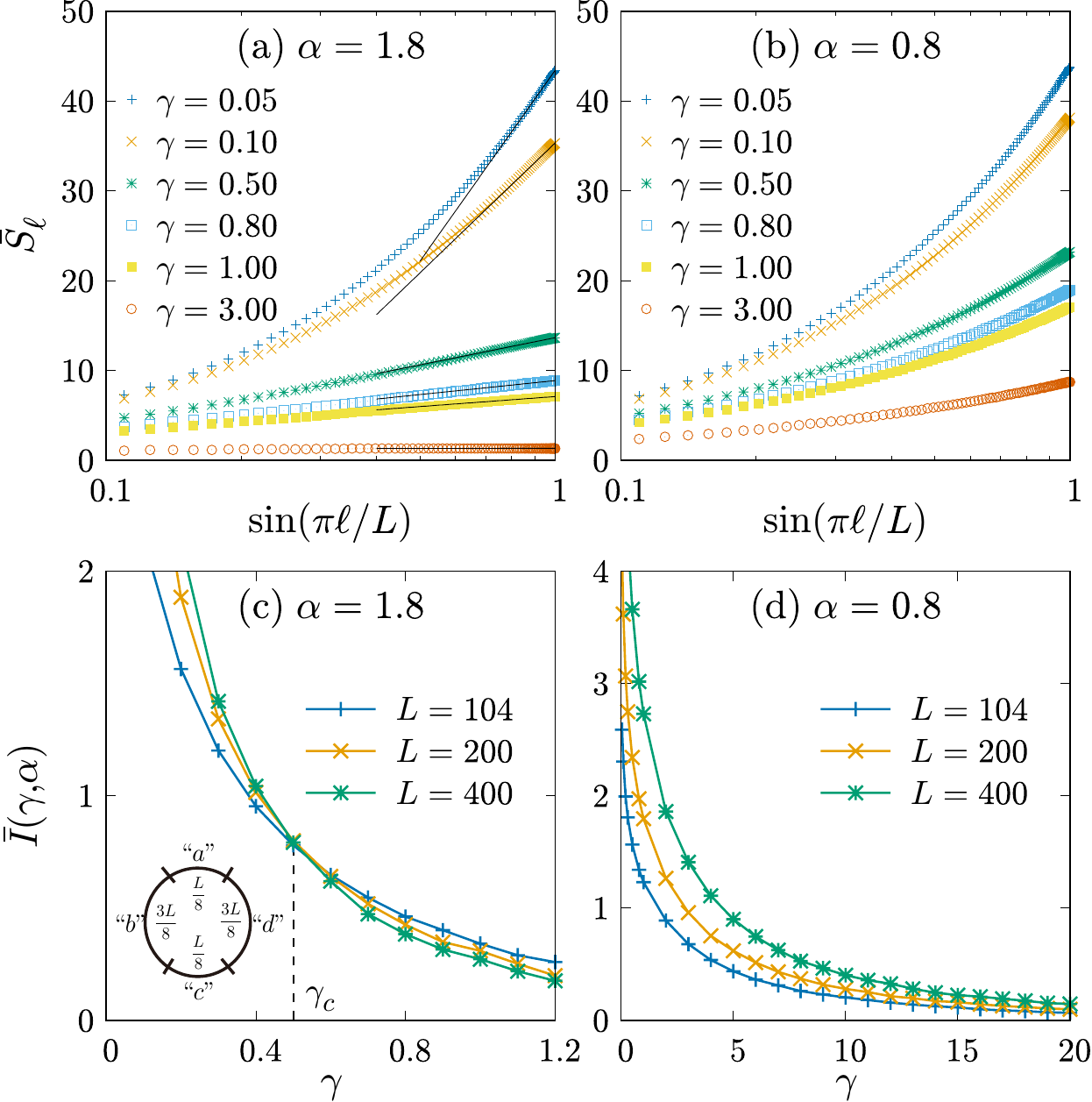}
  \caption{ (a) and (b): $\bar{S}_{\ell}$ as a function of $\sin (\pi \ell / L)$~$(L=200)$ for $\alpha=1.8$ in (a) and for $\alpha=0.8$ in (b), where the x-axis is log scale. The CFT behavior can be observed around $\ell\sim L/2$ as indicated in (a), where the solid lines are guides for the eyes.
   (c) and (d): The mutual information as a function of $\gamma$. }
  \label{fig2}
  }
\end{figure}

\begin{figure}[t]
\centering
{
  \includegraphics[clip, width=\linewidth]{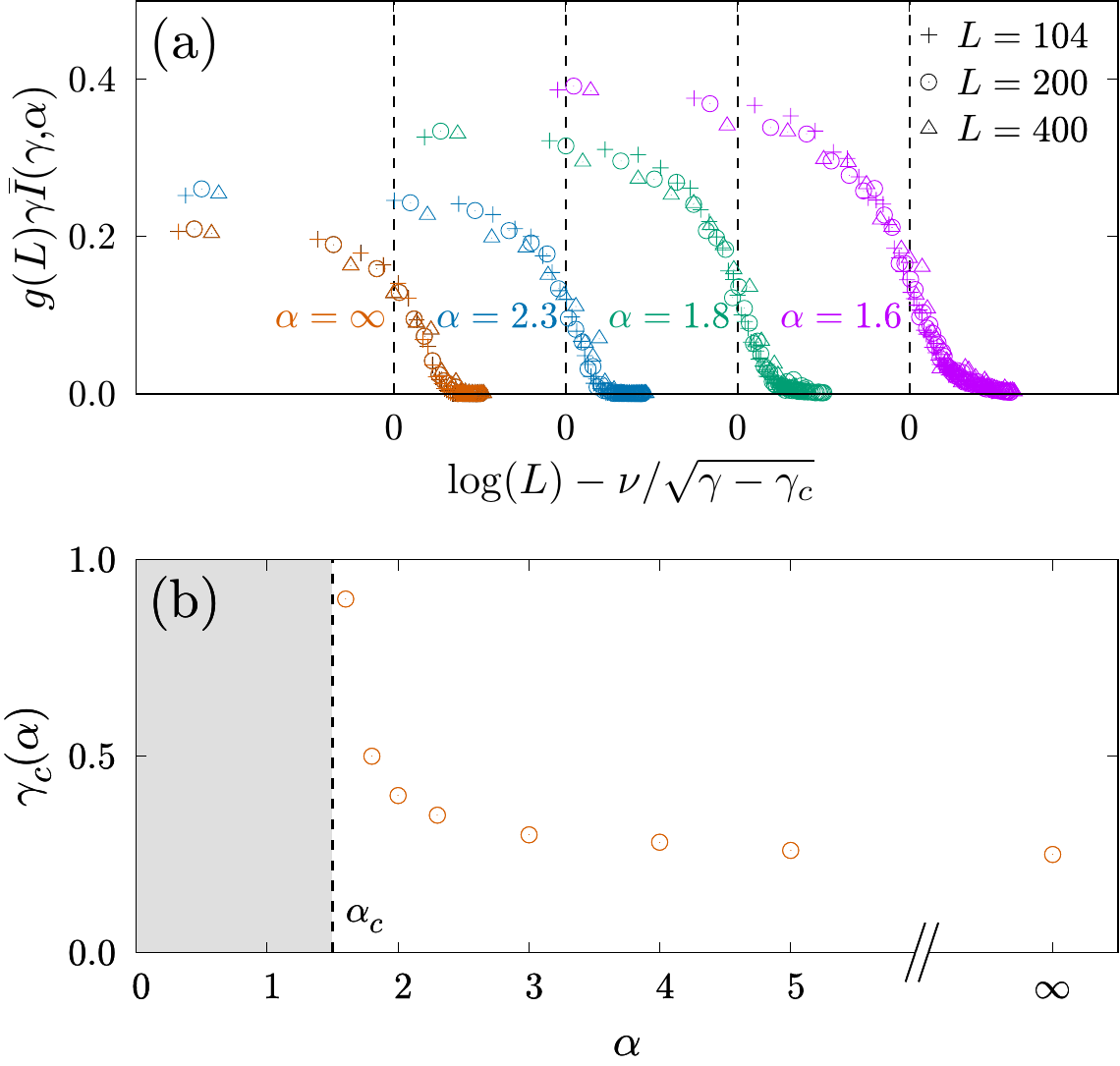}
  \caption{(a): Finite-size scaling for the mutual information with the BKT scenario $ g(L)\gamma \bar{I} (\gamma , \alpha)$ versus $\log (L \xi (\gamma, \gamma_{c}(\alpha)))=\log L - \nu /\sqrt{\gamma - \gamma_c (\alpha)}$, where $\nu=6.0, 5.5, 4.5$, and $3.0$ for $\alpha=1.6, 1.8, 2.3$, and $\alpha=\infty$, respectively and $g(L)=[1+1/(2\log L - 4)]^{-1}$. (b): Critical amplitude $\gamma_c (\alpha)$ as a function of $\alpha$. No critical amplitude exists for $\alpha < 1.5$.}
  \label{fig3}
  }
\end{figure}

\sectionprl{Free fermion case}
We first consider the free fermion case, that is, $V=0$. The wave function can be expressed in the form of
$  |\psi (t) \rangle = \prod_{m=1}^{L/2} \sum_{j=1}^{L} u_{j,m}(t)\, c_{j}^{\dagger}\, |\vac \rangle $.
The many-body wave function is expressed through the function $u_{j,m}(t)$, where the relation $\sum_{j=1}^{L} u_{j,m}^{\ast} u_{j,m'} =\delta_{m,m'}$ is imposed to guarantee normalization. The correlation function is calculated using the relation $\langle \psi (t) | c_i^{\dagger} c_j | \psi (t) \rangle =\sum_{m=1}^{L/2} u_{j,m}(t) u_{i,m}^{\ast}(t)$. In addition, the entanglement entropy can be computed once the correlation functions are obtained \cite{eisert2010colloquium}. See the supplementary material (SM) for these established methods \cite{suppl}.

In Fig.~\ref{fig2} (a) and (b), we show the numerical data for the entanglement entropy, which is a long-time average starting from the Neel state. See footnote \cite{ft0}
for the numerical details used to obtain the data. These figures show the $\ell$-dependence of the entanglement entropy for $\alpha=1.8$ and $\alpha=0.8$, respectively, for a fixed length $L=200$.
As indicated in Fig.~\ref{fig2} (a) with solid lines, the entanglement entropy around $\ell \sim L/2$ is well fitted by the functional form of the conformal field theory (CFT) with the effective central charge $c (\gamma, \alpha)$:
$  \bar{S}_{\ell} = (c(\gamma, \alpha)/3) \log_2 \left[(L/\pi) \sin (\pi \ell / L) \right] + {\rm const} \,  \label{cftrel}$.
For the short-range interaction limit $\alpha\to\infty$, this behavior was reported in Ref.~\cite{
  alberton2021entanglement}. 
In Fig.~\ref{fig2} (a), the entanglement entropies becomes constants for large $\gamma$, implying the area law, while Fig.~\ref{fig2} (a) has no indication of the area law. Figs.~\ref{fig2} (c) and (d) show the mutual information $\bar{I}(\gamma , \alpha)$ between the regime $``a"$ and $``c"$ depicted schematically in Fig.~\ref{fig2} (c), where the regions $``a"$ and $``c"$ with the length $L/8$ are separated by the regime $``b"$ and $``d"$ with the length $3L/8$.
Fig.~\ref{fig2}~(c) is the result for $\alpha =1.8$, where there is a crossing point $\gamma_c$, while Fig.~\ref{fig2}~(d) for $\alpha =0.8$ does not exhibit such a crossing phenomenon.

For the free fermion model, the finite-size effect is significant \cite{alberton2021entanglement}, and hence it is not trivial to obtain critical points which separate the sub-volume law phase (non-zero central charge in $\bar{S}_{\ell}$) and the area law phase (constants in $\bar{S}_{L/2}$). To suppress the size-effects, we use crossing points as illustrated in Fig.~\ref{fig2}~(c) to detect the critical points. In addition, we consider the Berezinskii-Kosterlitz-Thouless (BKT) scenario that was valid for the short-range limit \cite{alberton2021entanglement}. We remark that a similar technique using crossing points has been employed for several equilibrium systems \cite{carrasquilla2012superfluid,nishino2015termination}. Note also that the mutual information has been employed as a good indicator to detect the MIP in many systems \cite{Li_2019,Szyniszewski_2019,fuji2020measurement}. We use an ansatz of the finite-size scaling for the BKT scenario \cite{sandvik2010computational,harada1997universal,alberton2021entanglement}: $g (L) \gamma \bar{I}(\gamma , \alpha) = F[\log (L \xi (\gamma,\gamma_c(\alpha)) )]$, where $\xi$ is the BKT-type correlation length $\xi (\gamma , \gamma_c (\alpha)) \sim \exp (-\nu / \sqrt{\gamma - \gamma_c(\alpha)}) $. Here, $\gamma_c(\alpha )$ is the crossing point, which depends on $\alpha$. In Fig.~\ref{fig3}~(a), we verified that the scaling ansatz. These scaling data strongly suggest that $\gamma_c$ can be regarded as critical points. In Fig.~\ref{fig3}~(b), we show the behavior of $\gamma_c (\alpha)$ as a function of $\alpha$. In the gray shaded area ($\alpha <1.5$), we can find no critical points (that is, no crossing point, as shown in Fig.~\ref{fig2}~(d) \cite{suppl}). Thus, we find that the critical value $\alpha_c \simeq 1.5$ separates the absence and existence of the MIP (see Fig.~\ref{fig1}).

\sectionprl{Analysis based on the entanglement growth rate}
Next, we discuss the underlying physical mechanism for the numerical findings in Fig.~\ref{fig3}~(b). We argue that the key component is the growth rate of the entanglement entropy in pure quantum dynamics without measurement. Note the following expression for the entropy growth rate under pure quantum dynamics:
\begin{align}
  \begin{split}
  \dot{S}_{\ell} &= -i \| H_{AB}\| \lambda (\rho ) \, , \label{sdot}\\
    \lambda (\rho ) &:= {\rm Tr} \left( h_{AB}[ \rho , \rho_A\otimes {\bm 1}_B]\right) \, ,
   \end{split}
\end{align}
where $\rho = |\psi (t) \rangle \langle \psi (t)|$, $h_{AB} := H_{AB}/ \| H_{AB} \|$ ($\| ...\|$ is the operator norm), and ${\bm 1}_B$ is the identity operator for subsystem $B$. The Hamiltonian $H_{AB}$ denotes the boundary interaction between subsystems $A$ and $B$:
\begin{align}
  H_{AB}&= \sum_{i\in A}\sum_{j \in B} h_{i,j} \, ,
\end{align}
where $h_{i,j}$ is an interaction operator acting on sites $i$ and $j$. Of interest is the case where $\ell=L/2$. While the function $\lambda (\rho)$ highly depends on the states \cite{van2013entanglement,bennett2003capacities,bravyi2007upper}, the value is finite for less entangled states (it is zero, especially for a decoupled state such as the Neel state). Suppose that $\| H_{AB}\| $ is finite. Then, a sufficiently large measurement amplitude can suppress the entanglement growth, and hence the MIP should occur. Conversely, when $\| H_{AB}\| $ diverges in the thermodynamic limit, the finite measurement amplitude can no longer suppress entanglement growth, leading to the absence of the MIP. Therefore, the operator norm $\| H_{AB}\| $ should play a central role in determining the presence or absence of MIPs. We consider the size dependence of the operator norm $\| H_{AB}\|$ for $\ell=L/2$ in the free fermion case.  We numerically find that
$\| H_{AB}\| \propto L^{1- \alpha} ~(\alpha <1),~\propto \log L ~(1<\alpha < 1.5)$ and constants $(\alpha >1.5)$ (see also Fig.~S7 in SM \cite{suppl}).
 This explains the absence of MIP for $\alpha <1.5$ in Fig.~\ref{fig3}~(b).

\begin{figure*}[bt]
\centering
{
  \includegraphics[clip, width=\linewidth]{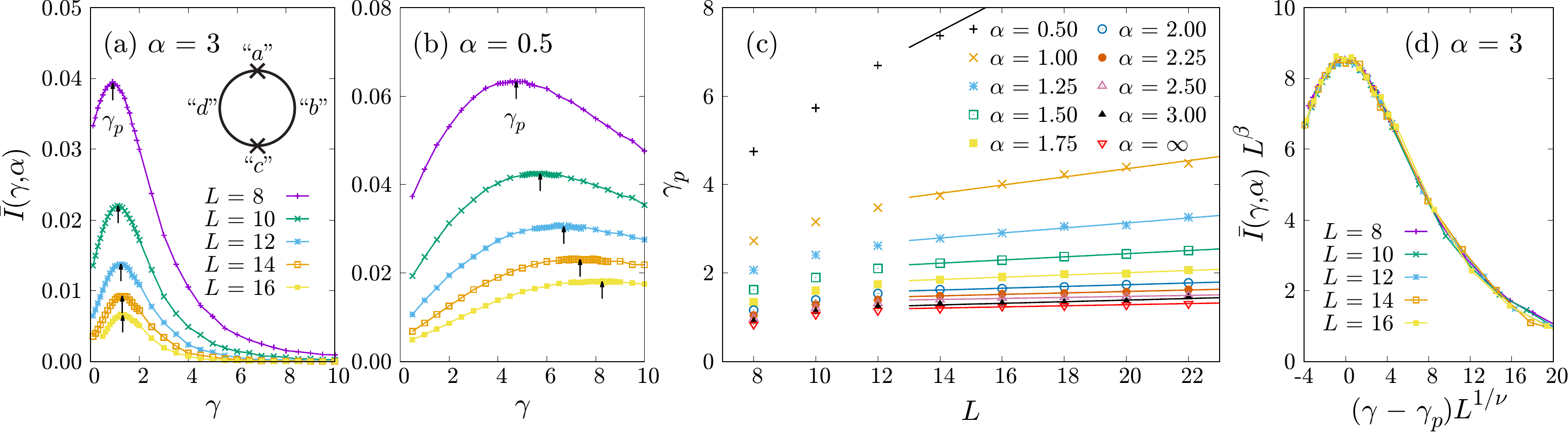}
  \caption{(a) and (b): The mutual information $\bar{I}(\gamma, \alpha)$ between the farthest two sites for $\alpha=3.0$ in (a) and for $\alpha=0.5$ in (b). The values $\gamma_p$ indicated by arrows are amplitudes that give the peaks. (c): The system-size dependence of $\gamma_p$.
(d): The finite-size scaling for $\alpha=3.0$ with the ansatz $\bar{I}(\gamma, \alpha )=L^{-\beta}f((\gamma - \gamma_p)L^{1/\nu})$ with the exponents $\beta = 2.59 \pm 0.02$ and $\nu = 1.4 \pm 0.1$.
  }
  \label{fig4}
  }
\end{figure*}
 
\sectionprl{Sufficient condition for the MIP in generic systems}
The free fermion model indicates that the behavior of the boundary interaction Hamiltonian is a key component for observing the MIP, as it governs the entanglement growth rate under pure quantum dynamics, and we now use this key component to make a statement applicable to generic systems. We discuss the sufficient conditions to observe the MIP for generic fermion systems. That is, we seek the value $\alpha_{sc}$ ($\ge \alpha_c$), where for $\alpha > \alpha_{sc}$, the MIP exists for generic many-body quantum systems. To this end, we rigorously derive the following lemma:

\begin{lemma} \label{suff_theorem}
Let us consider a Hamiltonian on the $d$-dimensional hypercubic lattice as $H=\sum_{i<j} h_{i,j}$ such that $\| h_{i,j} \| \le g /r^{\alpha}_{i,j}$, where $g$ is a constant.
Then, under the condition of $\alpha>\alpha_{sc}$ with
\begin{eqnarray}
    \alpha_{sc} &=& \left\{
  \begin{tabular}{ll}
     ${d/ 2}+1$  & ~~~{\it Bilinear\,  Systems}, \\
     $d+1$  & ~~~{\it Interacting\,  Systems},
  \end{tabular}
    \right. \label{alphasc}
  \end{eqnarray}
the operator norm $\| H_{AB}\|$ is upper-bounded by the boundary area between $A$ and $B$.
\end{lemma}
See the SM for rigorous proof \cite{suppl}. Using this lemma and the underlying physics, we can make a physical statement regarding the existence of the MIP:
\begin{statement} \label{cond_MIP}
  {\it $\alpha > \alpha_{sc}$ is a sufficient condition for the existence of the MIP for generic uniform and non-commuting systems.}
\end{statement}
Note that the statement is the sufficient condition for any system to have the MIP, and hence, this does not exclude that specific models can show the MIP for $\alpha <\alpha_{sc}$.
True critical value for a specific model, $\alpha_c$, is equal to or smaller than the value $\alpha_{sc}$. Intriguingly, in the bilinear fermion system that we have discussed, the above statement is optimal, since $\alpha_{c} =\alpha_{sc}=3/2$.

\sectionprl{Interacting fermion systems}
In the remainder of this paper, we demonstrate that the above statement is satisfied in non-integrable models. We consider the interacting fermion system with $V=1$ in (\ref{hamil}). We first calculate the operator norm $\| H_{AB} \|$ up to the size $L=256$ using the density-matrix renormalization group technique \cite{White1992, White1993, Schollwock2011}. We find clear evidence that $\| H_{AB} \| \propto L^{2-\alpha} $ for $\alpha < 2$, while they are constants for $\alpha >2$ (see Fig.~S8 in SM \cite{suppl}), from which we anticipate that the MIP appears for $\alpha >2$, which is consistent with Statement~\ref{cond_MIP}.

To assess this statement in greater detail, we perform the time-evolution calculation up to the system size $L=22$ using
the time-dependent variational principle method \cite{Haegeman2011, Haegeman2013, Yang2020}. See the footnote for the numerical details \cite{ft1}.
We focus on the mutual information $\bar{I}(\gamma,\alpha)$ between between the farthest two sites $``a"$ and $``c"$ that is depicted in Fig.\ref{fig4}~(a), where the regimes $``b"$ and $``d"$ have $L/2-1$ sites.

We show the results for $\alpha=3.0$ and $\alpha=0.5$ in Figs.~\ref{fig4}~(a) and (b), respectively, as functions of $\gamma$ for different system sizes. To discuss the figures, we recall that many studies so far \cite{Li_2019,Szyniszewski_2019,fuji2020measurement} have established that in non-integrable systems, the mutual information shows a peak as a function of $\gamma$, where the amplitude giving a peak denoted by $\gamma_p$ is identified as a critical value of the measurement amplitude separating the volume law phase for $\gamma < \gamma_p$, and the area law phase for $\gamma > \gamma_p$ in the thermodynamic limit. Figs.~\ref{fig4}(a) and (b) also show the peak structure as a function of $\gamma$. However, a crucial observation here is that the values of $\gamma_p$ (indicated by arrows) generally depend on the system size. In the case of $\alpha=3.0$, the values of $\gamma_p$ are not affected by the system size, especially for large $L$. For $\alpha =0.5$, the values of $\gamma_p$ are strongly affected by the system size, that is, $\gamma_p$ systematically increases with increasing system size.

This systematic change for $\alpha=0.5$ indicates that $\gamma_p$ eventually diverges in the thermodynamic limit, leading to the absence of the MIP. From this observation, the system-size dependence of $\gamma_p$ should be an indicator of the existence of the MIP. In Fig.~\ref{fig4}~(c), we plot the values of $\gamma_p$ as a function of the system size $L$ for various $\alpha$ values. This figure shows that for $\alpha >2$, $\gamma_p$ is robustly finite for sufficiently large system sizes, which means that the critical measurement amplitude exists even in the thermodynamic limit; hence, the MIP shows up for (at least) $\alpha > 2$. This observation is consistent with Statement~\ref{cond_MIP}, which states that $\alpha >2$ is sufficient to observe the MIP in one-dimensional generic interacting systems.
As an additional check on the existence of the MIP for $\alpha >2$, we consider the finite-size scaling with the ansatz $\bar{I}(\gamma, \alpha )=L^{-\beta}f((\gamma - \gamma_p)L^{1/\nu})$. In Fig. \ref{fig4}~(d), we show that the finite-size scaling works well with the exponents $\beta = 2.59 \pm 0.02$ and $\nu = 1.4 \pm 0.1$. Note that for $\alpha < 2$, this scaling is not available since $\gamma_p$ varies as increasing the size.
  Available analysis for given data are consistent with Statement~\ref{cond_MIP}. In the present interacting system, the sufficient condition is optimal, since $\alpha_{sc}=\alpha_c=2$.

\sectionprl{Summary}
We have revealed the effects of long-range interactions on the measurement-induced phase transition (MIP), which is summarized in Fig.~\ref{fig1}.
The key component for the existence of the MIP is the boundary interaction Hamiltonian under pure quantum dynamics in the thermodynamic limit. Based on this, we have arrived at sufficient conditions to observe the MIP, as described in Statement~\ref{cond_MIP}. The numerical results for the specific models indicate that this condition is optimal.
  We hope that this criterion is useful in real experimental setup with long-range interaction ~\cite{bendkowsky2009observation,bloch2008many,saffman2010quantum,yan2013observation,aikawa2012bose,britton2012engineered,islam2013emergence,zeiher2016many,zeiher2017coherent,bernien2017probing,zhang2017observation,neyenhuis2017observation,liu2019confined}.

\bigskip

\section*{Acknowledgement}
We are grateful to Yohei Fuji, Michael Buchhold, and Sebastian Diehl for providing details regarding their papers and useful suggestions. We also thank Seiji Miyashita for his useful comments on the BKT scaling. K.Su. was supported by Grants-in-Aid for Scientific Research (JP19K14644, JP20H01849).T.K. was supported by the RIKEN Center for AIP and JSPS KAKENHI (Grant No. 18K13475). K.Sa. was supported by Grants-in-Aid for Scientific Research (JP19H05603, JP19H05791).

{\it Note added}: After completing this work, we became aware of complementary works on the measurement-induced phase transition using long-range quantum circuits~\cite{block2021} and on the field theoretical argument for the free fermion systems~\cite{muller2021}. The latter work \cite{muller2021} is related to our sufficient condition in Statement~\ref{cond_MIP} and some classification of phase transitions is shown. 

\bibliography{MIP2.bib}

\begin{thebibliography}{73}
\bibitem [{ft0()}]{ft0}%
  \BibitemOpen
\bibfield  {journal} {  }\href@noop {} {}Numerical details: We set the
  relaxation time to $t=30$. We use $10^5$ trajectories for averaging. For each
  trajectory, we shift the coordinate of sites along the ring to take more
  average. Totally, we use $L\times 10^5$ samples to obtain
  $\bar{S}$.\BibitemShut {Stop}%
\end{thebibliography}

\clearpage

\pagestyle{empty}

\makeatletter
\long\def\@makecaption#1#2{{
\advance\leftskip1cm
\advance\rightskip1cm
\vskip\abovecaptionskip
\sbox\@tempboxa{#1: #2}%
\ifdim \wd\@tempboxa >\hsize
 #1: #2\par
\else
\global \@minipagefalse
\hb@xt@\hsize{\hfil\box\@tempboxa\hfil}%
\fi
\vskip\belowcaptionskip}}
\makeatother

\setcounter{equation}{0}
\def\theequation{A.\arabic{equation}}

\begin{widetext}

\begin{center}
{\large \bf Supplemental Material for \protect \\
  ``Fate of measurement-induced phase transition in long-range interactions'' }\\
\vspace*{0.3cm}
Takaaki Minato$^{1}$, Koudai Sugimoto$^{1}$, Tomotaka Kuwahara$^{2}$, and  Keiji Saito$^{1}$
\\
\vspace*{0.1cm}

$^{1}${\small \it Department of Physics, Keio University, Hiyoshi, Kohoku-ku, Yokohama 223-8522, Japan} \\

$^{2}${\small \it Mathematical Science Team, RIKEN Center for Advanced Intelligence Project (AIP), 1-4-1 Nihonbashi, Chuo-ku, Tokyo 103-0027, Japan}\\

\end{center}

\setcounter{equation}{0}
\renewcommand{\theequation}{S.\arabic{equation}}
\renewcommand{\thefigure}{S\arabic{figure}}
\renewcommand{\bibnumfmt}[1]{[S#1]}

\section{Proof of the lemma 1}
We rigorously show the lemma in the main text, which is restated as follows
\begin{flushleft}
  {\bf Lemma 1}~{\it
Let us consider a Hamiltonian on a $d$-dimensional hypercubic lattice as $H=\sum_{i<j} h_{i,j}$ such that $\| h_{i,j} \| \le g /r^{\alpha}_{i,j}$, where $g$ is a constant.
Then, under the condition of $\alpha>\alpha_{sc}$ with
\begin{eqnarray}
    \alpha_{sc} &=& \left\{
  \begin{tabular}{ll}
     ${d/ 2}+1$  & ~~~{\it Bilinear\,  Systems}, \\
     $d+1$  & ~~~{\it Interacting\,  Systems},
  \end{tabular}
    \right. \label{suppl.lemma}
  \end{eqnarray}
the operator norm $\| H_{AB}\|$ is upper-bounded by the boundary area between $A$ and $B$.
  }
\end{flushleft}
Below, we present the proof of the lemma. We consider a $d$-dimensional hypercubic lattice with linear size $L$ and unit spacing. We impose an open boundary condition in the direction perpendicular to the boundary surface. For $d\ge 2$, we additionally impose a periodic boundary condition in the direction parallel to the boundary surface. We can write the operator norm of the boundary interaction Hamiltonian per boundary surface area ${\cal A}$ as follows:
\begin{align}
\| H_{AB} \|/{\cal A}  &:=  \| \sum_{i \in A} \sum_{j \in B} h_{i,j} \| /{\cal A}\le \sum_{i \in A} \| \sum_{j \in B} h_{i,j}\| /{\cal A}   \, , \label{suppl.lemma.proof1}
\end{align}
where ${\cal A}$ is the boundary surface area between $A$ and $B$, that is, ${\cal A}=L^{d-1}$.
We consider the r.h.s. in (\ref{suppl.lemma.proof1}). We first consider the case of bilinear systems. In the bilinear Hamiltonian case, the local Hamiltonian can be generically written as:
\begin{align}
  h_{i,j} &= (g_{i,j} c_{i}^{\dagger}c_{j} + g_{i,j}' c_{i} c_{j} + h.c.)/r_{i,j}^{\alpha} ,
\end{align}
where $g_{i,j}$ and $g_{i,j}'$ are $O(1)$. Then, we estimate the bound in (\ref{suppl.lemma.proof1}). To this end, we define
\begin{align}
  \begin{split}
  d_{i}& :={\mu_i}^{-1} \sum_{j\in B} g_{i,j} c_j / r_{i,j}^{\alpha} \, , ~~~ d_{i}':={\mu_i'}^{-1} \sum_{j\in B} g_{i,j}' c_j / r_{i,j}^{\alpha}, , \\
  \mu_i & :=\left[ \sum_{j\in B} |g_{i,j}|^2/r_{i,j}^{2\alpha} \right]^{1/2},~ \mu_i':=\left[ \sum_{j\in B} |g_{i,j}'|^2/r_{i,j}^{2\alpha} \right]^{1/2} \, .
  \end{split}\label{suppl.bil.trick}
\end{align}
Then, we can estimate the bound in (\ref{suppl.lemma.proof1}) as follows
\begin{align}
  \sum_{i \in A} \| \sum_{j \in B} h_{i,j}\| /{\cal A}  &\le 2 {\cal A}^{-1}                                                         \sum_{i\in A}  \left[  \| c_i^{\dagger} \sum_{j\in B} g_{i,j} c_j /r_{i,j}^{\alpha}\|  + \|  c_i \sum_{j\in B} g_{i,j}'  c_j /r_{i,j}^{\alpha} \| \right] \nonumber \\
                                                        &= 2 {\cal A}^{-1} \sum_{i\in A}  \left[  \| c_i^{\dagger} \mu_i d_i \|  + \|  c_i \mu_i ' d_i' \| \right]\le 2{\cal A}^{-1} \sum_{i\in A} \mu_i + \mu_i' \le 4 {\cal A}^{-1}g_{\rm max}                                                          \sum_{i\in A} \left[ \sum_{j\in  B}1/r_{i,j}^{2\alpha }\right]^{1/2} \nonumber \\
                                                        &=4 {\cal A}^{-1}g_{\rm max} \sum_{x=1}^{L/2}\sum_{i:{\rm dist}(i:B)=x}
                                                          \left[ \sum_{y=1}^{L/2} \sum_{j:{\rm dist}(A:j)=y} 1/r_{i,j}^{2\alpha }\right]^{1/2} \nonumber \\
                                                        &=4 g_{\rm max}                                                         \sum_{x=1}^{L/2} \left[ \sum_{y=1}^{L/2} \sum_{j:{\rm dist}(A:j)=y} 1/r_{i_0,j}^{2\alpha }\right]^{1/2} ,
                                                          \label{suppl:bil2}
\end{align}
where $g_{\rm max}=\max (\{ g_{i,j} , g_{i,j}'\})$, ${\rm dist}(i:B):=\min_{j\in B} r_{i,j}$, and ${\rm dist}(A:j):=\min_{i\in A} r_{i,j}$. The site $i_0$ is some point in the regime $A$ satisfying ${\rm dist}(i_0:B)=x$.
We have used ${\cal A}=\sum_{i:{\rm dist}(i:B)=x}=L^{d-1}$, irrespective of $x$. 
Finally, we consider the r.h.s. in (\ref{suppl:bil2}). For $d=1$, we can estimate the bound as follows:
\begin{align}
\sum_{x=1}^{L/2} \left[ \sum_{y=1}^{L/2} \sum_{j:{\rm dist}(A:j)=y} 1/r_{i_0,j}^{2\alpha }\right]^{1/2}
  &= \sum_{x=1}^{L/2} \left[ \sum_{y=1}^{L/2}  1/(x+y)^{2\alpha }\right]^{1/2}
  =   \left[ \sum_{y=1}^{L/2}  1/(1+y)^{2\alpha }\right]^{1/2}
    + \sum_{x=2}^{L/2} \left[ \sum_{y=1}^{L/2}  1/(x+y)^{2\alpha }\right]^{1/2} \nonumber \\
  & \le (2\alpha -1)^{-1/2} + (2\alpha -1)^{-1/2}(3/2 - \alpha)^{-1}
    ((L/2)^{3/2 -\alpha} -1 ) \, ,
\end{align}
where in advance, we assumed $\alpha >1/2$ to obtain the final inequality. For $d\ge 2$, we can calculate the r.h.s. in (\ref{suppl:bil2}) as follows:
\begin{align}
  \sum_{x=1}^{L/2} \left[ \sum_{y=1}^{L/2} \sum_{j:{\rm dist}(A:j)=y}
  1/r_{i_0,j}^{2\alpha }\right]^{1/2}  \!\!\!\!\!
  &= \sum_{x=1}^1 \left[ \sum_{y=1}^{L/2}
    \sum_{j:{\rm dist}(A:j)=y} 1/r_{i_0,j}^{2\alpha }\right]^{1/2}
  +  \sum_{x=2}^{L/2} \left[ \sum_{y=1}^{L/2}
    \sum_{j:{\rm dist}(A:j)=y} 1/r_{i_0,j}^{2\alpha }\right]^{1/2} \nonumber \\
  &\le
    \left[ \Gamma_d \sum_{y=1}^{L/2} \int_0^{\infty} d\xi \xi^{d-2}/\left[ (1+y)^2 + \xi^2 \right]^{\alpha }\right]^{1/2}  \!\!\!\!+
    \sum_{x=1}^{L/2}  \left[\Gamma_d \sum_{y=1}^{L/2} \int_0^{\infty} d\xi \xi^{d-2}/\left[ (x+y+1)^2 +
    \xi^2 \right]^{\alpha }\right]^{1/2} \nonumber \\
  & =
    \left[ (\Gamma_d B/2) \sum_{y=1}^{L/2} (y+1)^{d-1-2\alpha} \right]^{1/2}  + \sum_{x=1}^{L/2}  \left[(\Gamma_d B/2) \sum_{y=1}^{L/2} (x+y+1)^{d-1-2\alpha} \right]^{1/2} \nonumber \\
  & \le
\left[ (\Gamma_d B/2)/(2\alpha - d) \right]^{1/2}
    + \left[ (\Gamma_d B/2)/(2\alpha - d) \right]^{1/2}
   \left[(L/2)^{-\alpha + d/2 +1} -1\right]/(-\alpha +d/2 +1) \, ,
\end{align}
where $\Gamma_d =2$ for $d=2$ and $2\pi$ for $d=3$, which is the coefficient of the $(d-1)$-dimensional spherical surface area, and $B$ is the beta function, $B(-1/2 + d/2, \alpha + 1/2 -d/2)$. In the calculation, we assumed $2\alpha - d >0$ in advance. Thus, the bilinear case in (\ref{suppl.lemma}) is proven.

In generic Hamiltonian systems, we cannot use the trick (\ref{suppl.bil.trick}); hence, we calculate the bound as follows:
\begin{align}
  \sum_{i \in A} \| \sum_{j \in B} h_{i,j}\| /{\cal A}
  &\le g{\cal A}^{-1} \sum_{i \in A} \sum_{j \in B} 1/r_{i,j}^{\alpha}
  \le g \sum_{x=1}^{L/2} \sum_{y=1}^{L/2} \sum_{j:{\rm dist}(A:j)=y} 1/r_{i_0,j}^{\alpha } \, .
\end{align}
For the $d=1$ case, we have
\begin{align}
  \sum_{x=1}^{L/2} \sum_{y=1}^{L/2} \sum_{j:{\rm dist}(A:j)=y} 1/r_{i_0,j}^{\alpha }
  &= \sum_{x=1}^{L/2} \sum_{y=1}^{L/2} 1/(x+y)^{\alpha }  \le 1/(\alpha -1) +
        \left[(L/2)^{-\alpha +2} -1\right]/\left[(\alpha -1)(-\alpha +2) \right] \, ,
\end{align}
where, in advance, we have assumed $\alpha -1 >0$. For $d\ge 2$, the bound is computed as follows:
\begin{align}
  \sum_{x=1}^{L/2} \sum_{y=1}^{L/2} \sum_{j:{\rm dist}(A:j)=y} 1/r_{i_0,j}^{\alpha }
  &= \sum_{x=1}^1 \sum_{y=1}^{L/2}
    \sum_{j:{\rm dist}(A:j)=y} 1/r_{i_0,j}^{\alpha }
  +  \sum_{x=2}^{L/2} \sum_{y=1}^{L/2}
    \sum_{j:{\rm dist}(A:j)=y} 1/r_{i_0,j}^{\alpha } \nonumber \\
  &\le
    \Gamma_d \sum_{y=1}^{L/2} \int_0^{\infty} d\xi \xi^{d-2}/\left[ (1+y)^2 + \xi^2 \right]^{\alpha /2}  + \sum_{x=1}^{L/2} \Gamma_d \sum_{y=1}^{L/2} \int_0^{\infty} d\xi \xi^{d-2}/\left[ (x+y+1)^2 + \xi^2 \right]^{\alpha /2} \nonumber \\
  &\le \Gamma_d B'/ \left[ 2(\alpha -d) \right] + \Gamma_d B'/ \left[ 2(\alpha -d) \right]
    \left[ (L/2)^{-\alpha + d+ 1} - 1\right]/(-\alpha + d +1) \, ,
\end{align}
where $\Gamma_d$ is the same as in the bilinear case, and $B'$ is the beta function $B'=B((d-1)/2, (\alpha - d + 1)/2)$. In the calculation, we assumed $\alpha -d >0$ in advance. Thus, the condition for generic systems in (\ref{suppl.lemma}) is proven.

\section{Numerical method for free fermion systems}

Let $L$ be the size of a one-dimensional lattice, and let $N$ be the total number of fermions. We consider the free fermion system as:
\begin{align}
  H &= \sum_{j=1}^L \sum_{r=1}^{L/2}{1\over r^{\alpha}} \left[ -c_{j+r}^{\dagger}c_j
      -c_{j}^{\dagger}c_{j+r} \right] \, .\label{suppl:hamil}
\end{align}
Because the Hamiltonian is bilinear, the many-body wave function is always cast in the following form:
\begin{align}
  |\psi (t) \rangle &= \prod_{m=1}^{N} \sum_{j=1}^{L} u_{j,m}(t)\, c_{j}^{\dagger}\, |\vac \rangle ,
\end{align}
where $|{\rm vac} \rangle$ is the vacuum state.
The relation $\sum_{j=1}^{L} u_{j,m}^{\ast} u_{j,m'} =\delta_{m,m'}$ guarantees the normalization $\langle \psi (t) | \psi(t) \rangle =1$ because $b_m^{\dagger}:=\sum_{j=1}^{L} u_{j,m} c_j^{\dagger}$ defines a new fermion operator satisfying the anti-commutation relations $\left\{ b_m, b_{m'}^{\dagger} \right\} =\delta_{m,m'}$ and $\left\{ b_m, b_{m'} \right\} =0$

The correlation function $\langle c_i^{\dagger} c_{i'} \rangle:= \langle \psi (t) |c_i^{\dagger} c_{i'} | \psi (t) \rangle$ is computed using the $L\times N$ matrix ${\bm  u}$. Note that $\langle c_i^{\dagger} c_{i'} \rangle =\delta_{i,i'} - \langle \psi (t) | c_{i'} c_i^{\dagger} | \psi (t) \rangle $, and $b_0 ' =c_{i'}$ and $b_0^{\dagger} = c_{i}^{\dagger}$. Then, we have
  \begin{eqnarray}
  \langle c_i^{\dagger} c_{i'} \rangle &=&  \langle {\rm vac} | b_N b_{N-1} \cdots b_1 b_0' b_0^{\dagger} b_1^{\dagger} \cdots b_{N-1}^{\dagger} b_N^{\dagger} |{\rm vac} \rangle =\det {\bm A} \, , \\
  A_{i,j} &=& \left\{
            \begin{array}{ll}
              \langle {\rm vac}|  b_{0}' b_{j}^{\dagger}|{\rm vac}\rangle & ~~~i=0, ~\& ~ j=0,\cdots, N \\
              \langle {\rm vac}|  b_{i} b_{j}^{\dagger}|{\rm vac}\rangle & ~~~{\rm otherwise}
              \end{array}
            \right. ,
\end{eqnarray}
where ${\bm A}$ is the $(N+1)\times (N+1)$ matrix. Note that $\langle {\rm vac}|  b_{i} b_{j}^{\dagger}|{\rm vac}\rangle =\delta_{i,j}$ for $i,j=1,\cdots ,N$. Then, after simple algebra using elementary mathematical properties on the determinant, we obtain the relation
\begin{align}
 \langle c_i^{\dagger} c_{i'} \rangle &= \sum_{m=1}^{N} u_{i,m} u_{i' , m}^{\ast} \, .
\end{align}

The entanglement entropy is computed following the standard argument \cite{eisert2010colloquium}, once the correlation matrix is obtained. Suppose that we consider subsystem $A$ consisting of $\ell$ sites $j=1,\cdots , \ell$. Let $\rho_A$ be a reduced density matrix for subsystem $A$, which is generically written in the following Gaussian form:
\begin{align}
  \rho_{A} &= \exp \left( -\sum_{m=1}^{\ell} a_m d_{m}^{\dagger} d_m \right)/ \prod_{m=1}^{\ell} (1 + e^{-a_m}) \, ,
\end{align}
where $d_m$ is a fermion operator connected to the original fermion operator $\{ c_j \}$ through a unitary transform $d_m = \sum_{j=1}^{\ell} v_{m,j} c_{j}$. Then, the entanglement entropy $S_{\ell}$ is given as
\begin{align}
  S_{\ell} &= -\sum_{m=1}^{\ell} \bigl[ \langle d_m^{\dagger } d_m \rangle \log \langle d_m^{\dagger } d_m \rangle.
             + \langle d_m d_m^{\dagger } \rangle \log \langle d_m d_m^{\dagger }    \rangle
             \bigr] \, ,
\end{align}
where $\langle d_m^{\dagger } d_m \rangle$ can be computed by diagonalizing the correlation matrix ${\bm D}$, the element of which is given by $D_{i,j}= \langle c_i^{\dagger } c_j \rangle$ ~$(i,j=1,\cdots, \ell)$.

We explain the protocol for computing the time evolution of the wave function $\psi (t)$, that is, equivalently, the time evolution of the $L\times N$ matrix ${\bm u} (t) $ \cite{alberton2021entanglement}.
(i): We determine the jump time $\tau=-\log (r)/(\gamma N)$ with a uniformly distributed random number $r \in [0,1]$. (ii): During the time duration $\tau$, the unitary time evolution is performed with the one-particle Hamiltonian ${\bm h}$ for the matrix ${\bm u}$ as ${\bm u}(t + \tau) = e^{-i{\bm h} \tau} {\bm u} (t)$. (iii): Site $j$ is measured according to the probability $P_j = \langle \psi (t + \tau) | c_j^{\dagger} c_j | \psi (t + \tau ) \rangle/N $. (iv): The post-measurement state $|\psi '\rangle $ after the measurement of site $j$ is formally given as
\begin{align}
  |\psi ' \rangle &=
                  { c_j^{\dagger} c_j | \psi \rangle \over
                \sqrt{  \langle c_j^{\dagger} c_j \rangle }
                  } .
\end{align}
Using Wick's theorem, one finds the relation for the correlation function
\begin{eqnarray}
 D_{i,i'} := \langle \psi '  | c_i^{\dagger} c_{i'} | \psi ' \rangle  &=&
                                                             \left\{
                                                             \begin{array}{lll}
                                                               1 &~~~ i=i' = j \\
                                                               0 & ~~~(i=j, i' \neq i) ~{\rm or}~(i\neq i' , i' = j) \\
                                                               \langle c_i^{\dagger} c_{i'} \rangle - \langle c_j^{\dagger} c_{i'}\rangle  \langle c_i^{\dagger} c_{j}\rangle /\langle c_j^{\dagger} c_{j}\rangle & ~~~{\rm otherwise} \,
                                                             \end{array}
                                                             \right. .
\end{eqnarray}
The new matrix ${\bm u}$ is obtained through an SVD decomposition ${\bm D} = {\bm u} {\bm S} {\bm u}^{\dagger} $, where $S_{i,i}=1~(1\le i \le N)$ and $0 ~(N+1\le i \le L)$.

\section{Supplementary numerical data for free fermion systems}
Here, we present supplementary numerical data for the free fermion system. We first show the crossing behavior of the mutual information as a function of $\gamma$ in Fig.~\ref{fig12:suppl}. See the main text for the definition of mutual information. In the figures for $\alpha=5.0, 2.0$, and $1.6$, we clearly see that the data cross each other. However, for $\alpha=1.4, 1.3$ and $1.1$, curves overlap each other for the large $\gamma$ regime and hence, we cannot see the crossing phenomena. 

\begin{figure}[b]
  \begin{center}
    \begin{tabular}{c}
      \begin{minipage}{0.34\hsize}
        \begin{center}
                \includegraphics[width= \textwidth]{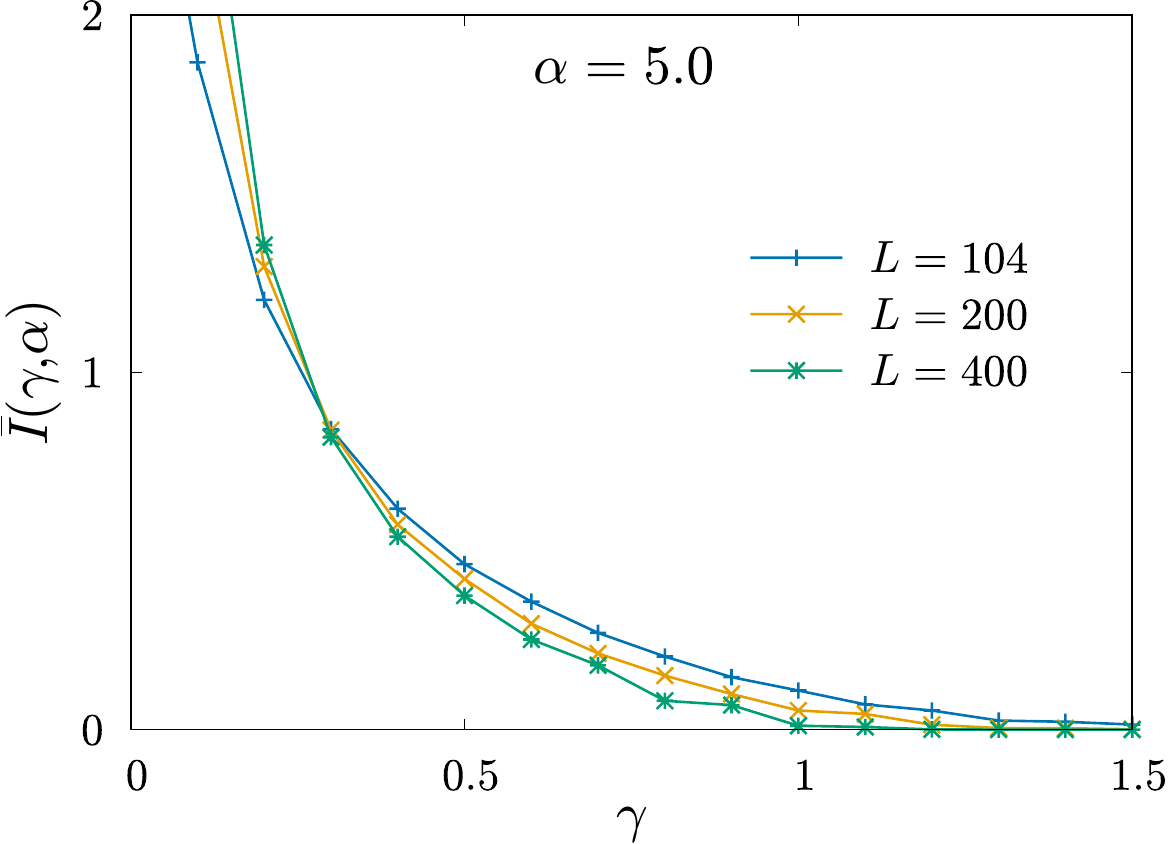}
        \end{center}
      \end{minipage}

      \begin{minipage}{0.34\hsize}
        \begin{center}
                \includegraphics[width= \textwidth]{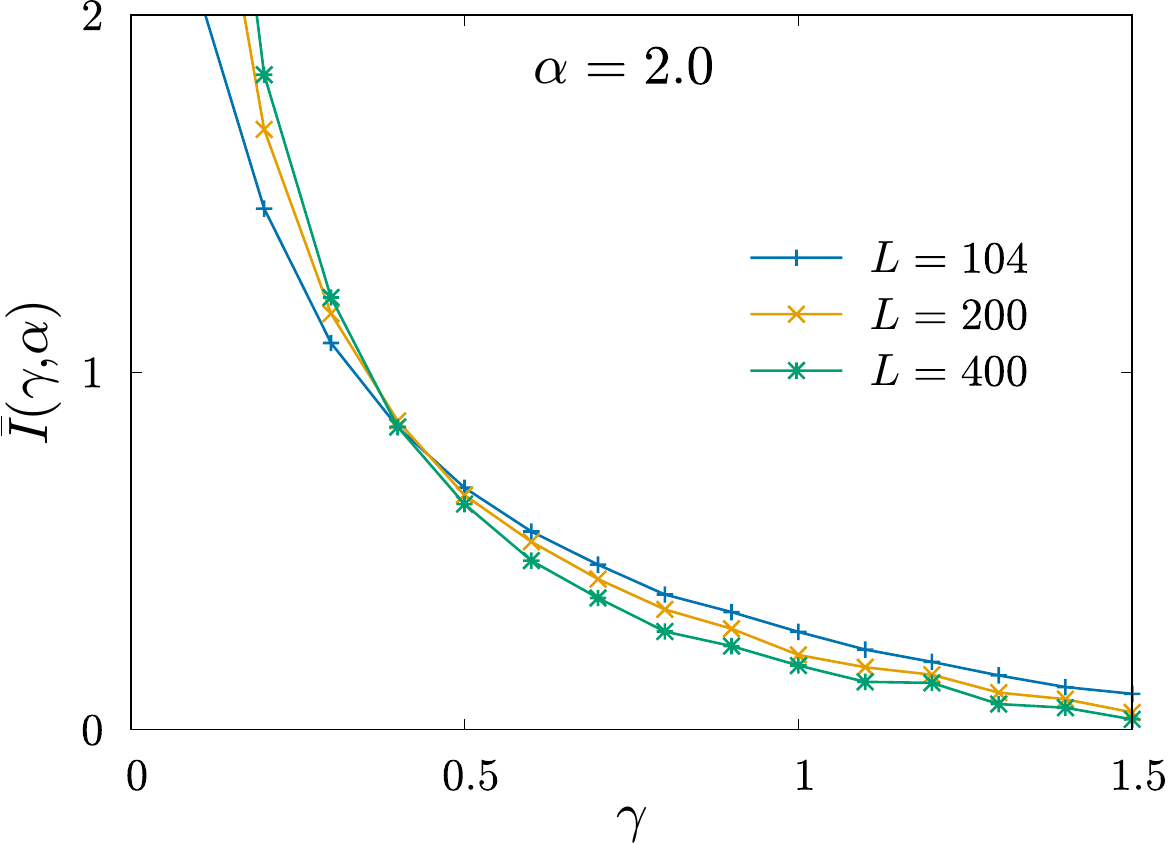}
        \end{center}
      \end{minipage}
      \begin{minipage}{0.34\hsize}
        \begin{center}
                \includegraphics[width= \textwidth]{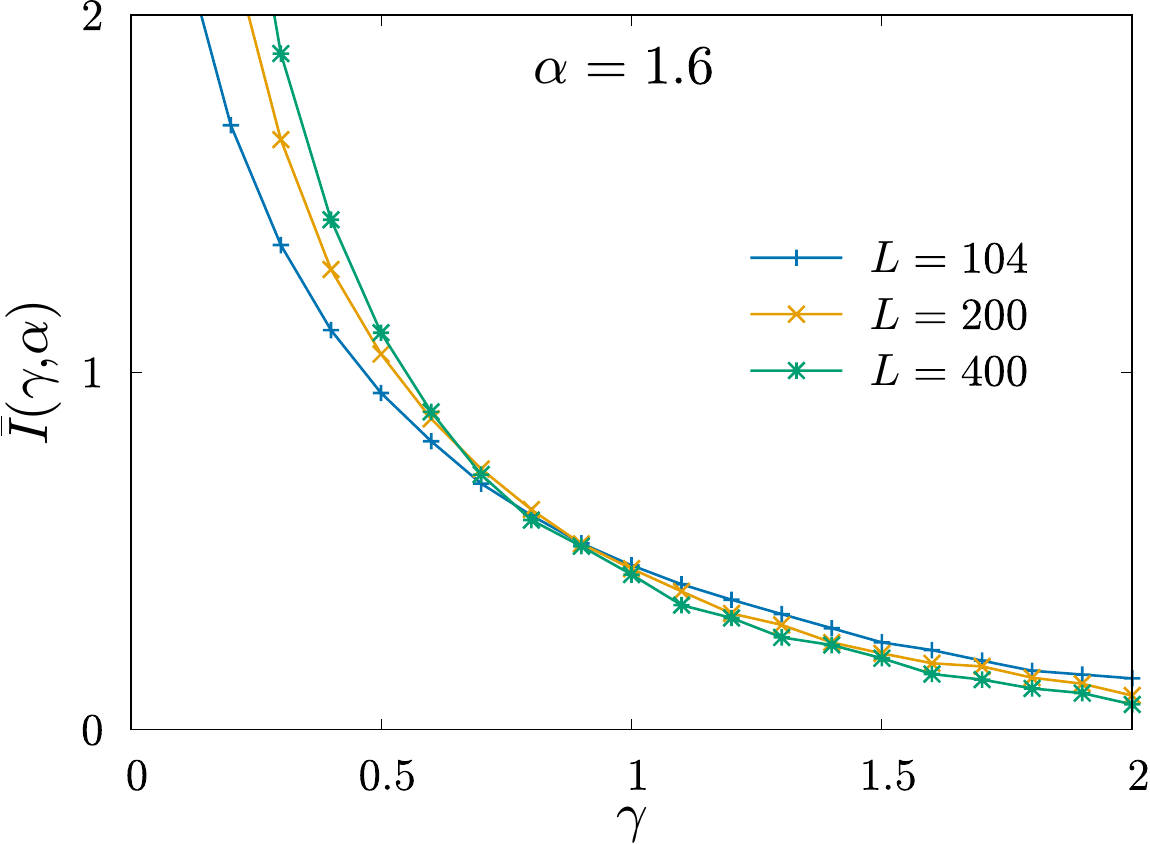}
        \end{center}
      \end{minipage}
    \end{tabular}
    \\
    \begin{tabular}{c}
      \begin{minipage}{0.34\hsize}
        \begin{center}
                \includegraphics[width= \textwidth]{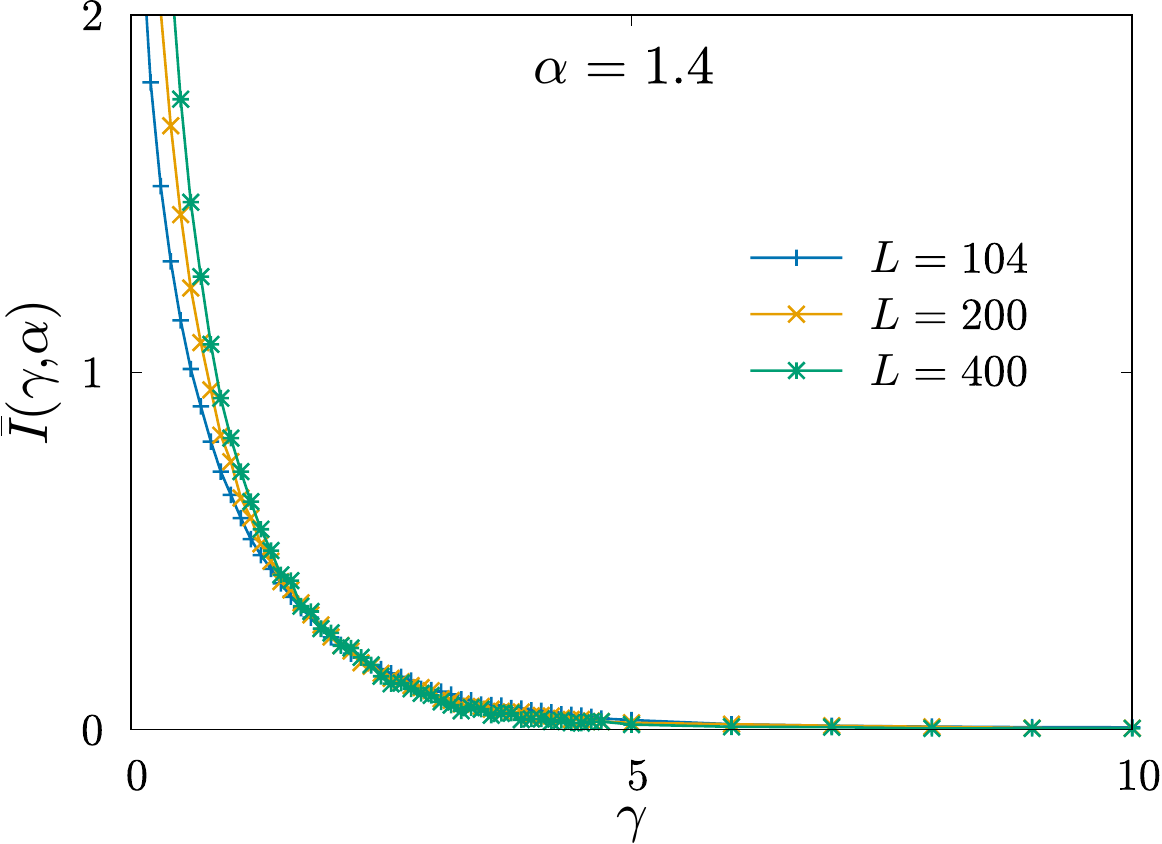}
        \end{center}
      \end{minipage}

      \begin{minipage}{0.34\hsize}
        \begin{center}
                \includegraphics[width= \textwidth]{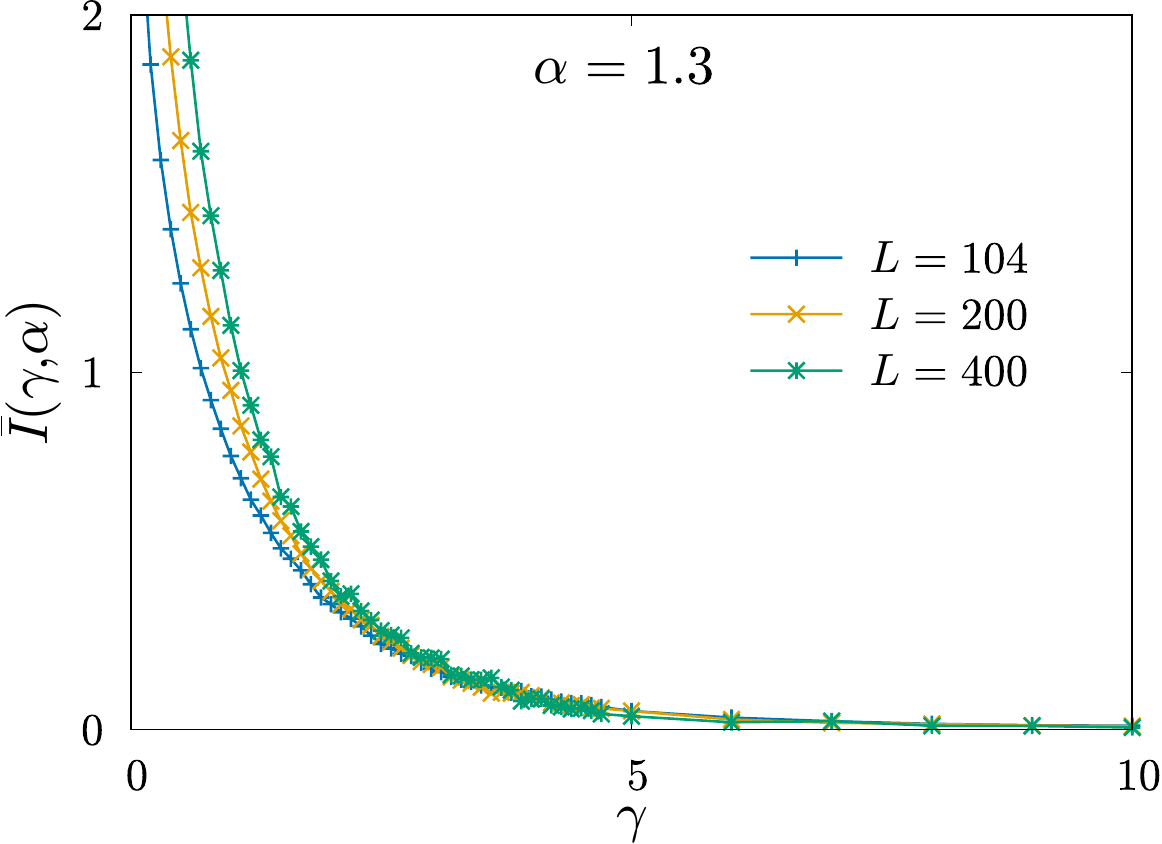}
        \end{center}
      \end{minipage}
      \begin{minipage}{0.34\hsize}
        \begin{center}
                \includegraphics[width= \textwidth]{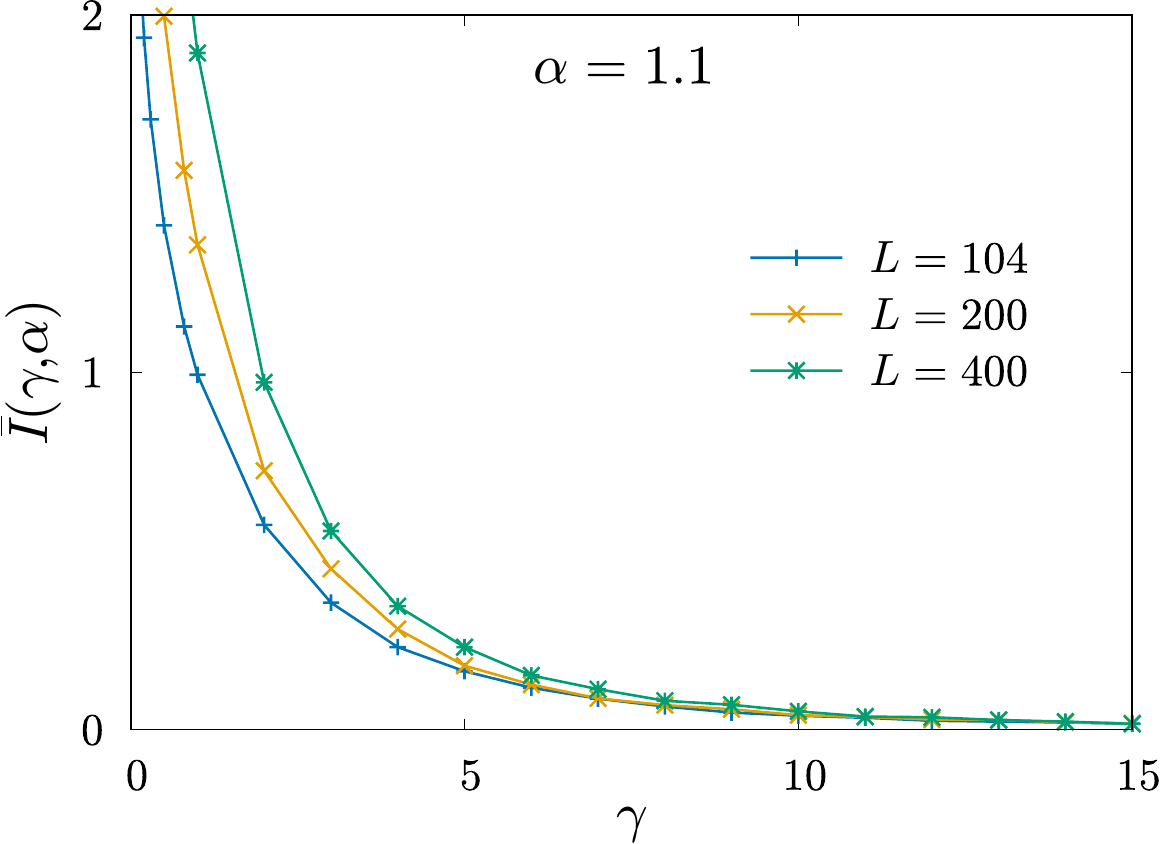}
        \end{center}
      \end{minipage}
    \end{tabular}
    
    \caption{\label{fig12:suppl}      Mutual informaations as a function of $\gamma$ for many cases of $\alpha$.}
  \end{center}
\end{figure}
In the main text, we showed the finite scaling analysis for the mutual information. We here show that the central charges show finite-size scaling. Getting the the mutual information data does not need any fitting procedures, and hence the mutual information is accurately computable.  The central charge is, however, obtained through the fitting procedure around $\ell=L/2$ for the entanglement entropies $\bar{S}_{\ell}$ as shown in Fig.1 (a) in the main text. Hence, we should have in mind that values of central charges are less accurate than the mutual information due to the fitting procedure.  We show the results of finite-size scaling for the central charges in Fig.~\ref{fig22:suppl}. Within available data, we  can see that the central charges are well scaled with the BKT scenario.
\begin{figure}[t]
\centering
\includegraphics[clip,width=0.5\linewidth]{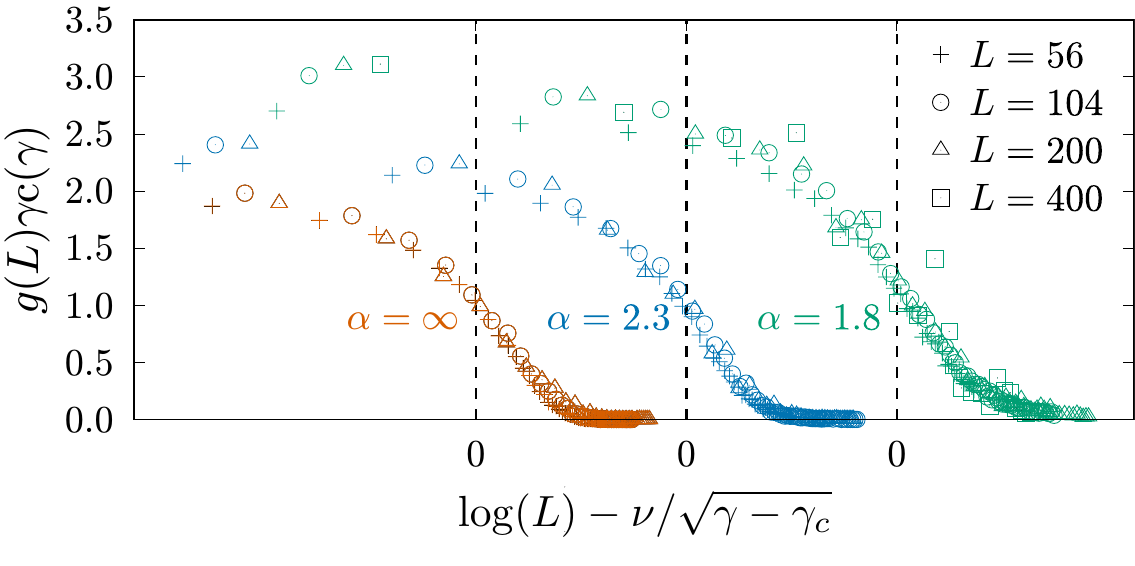}
  \caption{\label{fig22:suppl}The finite-size scaling for the central charges with the BKT scenario.
    $g(L)\gamma c(\gamma)$ versus $\log (L\xi(\gamma, \gamma_c (\alpha ) ) ) = \log L - \nu / \sqrt{\gamma - \gamma_c (\alpha) } $, where $\nu = 5.0, 4.3$ and $3.5$ for $\alpha=1.8, 2.3$ and $\infty$, respectively. 
  }
\end{figure}

Finally, we show the size dependence of the boundary interaction Hamiltonian $H_{AB}$, which is defined as
\begin{align}
  H_{AB}&:= \sum_{i\in A}\sum_{j \in B} h_{i,j} = \sum_{j=1}^{\ell} \left[\sum_{k=L/2+j}^L {1\over |L+j-k|^{\alpha}} + \sum_{k=\ell +1}^{j + L/2} {1\over |k-j|^{\alpha}}\right]   \left[ -c_j^{\dagger} c_k - c_k^{\dagger} c_j \right] ,
          \label{suppl:hamilab}
\end{align}
where we set the subsystem $A$ with length $L/2$. We numerically calculate the operator norm $\| H_{AB}\|$ as a function of the system size up to $L=25000$. The results are shown in Fig.~\ref{fig3:suppl}. The figures strongly suggest the following behavior:
\begin{eqnarray}
  \| H_{AB}\| &\propto
                \left\{
                \begin{array}{ll}
                  L^{1 - \alpha } & ~~\alpha < 1  \, , \\
                  \log L & ~~1< \alpha < 1.5  \, , \\
                  {\rm const.} &~~ \alpha > 1.5  \, .
                \end{array}
                \right. \label{suppl:habpowerfree}
\end{eqnarray}

\begin{figure}[]
    \begin{tabular}{c}
      \begin{minipage}{0.34\hsize}
        \begin{center}
                \includegraphics[width= \textwidth]{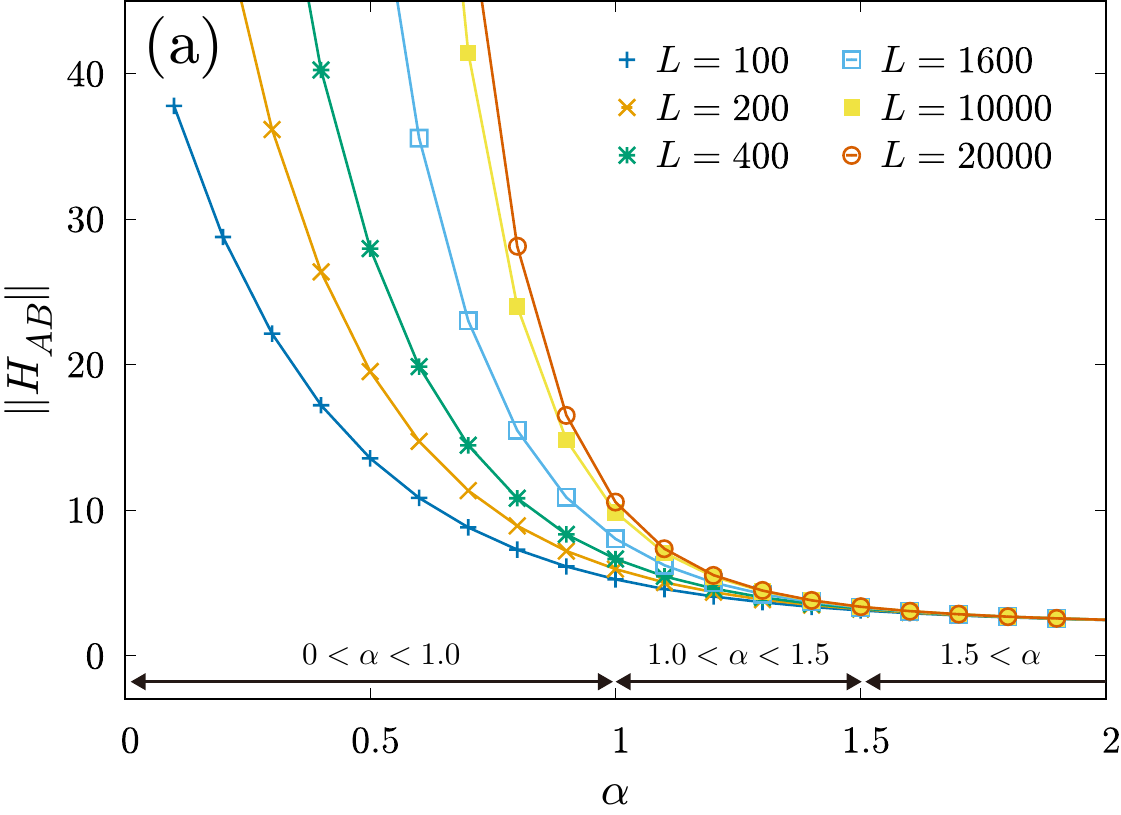}
        \end{center}
      \end{minipage}
      \begin{minipage}{0.355\hsize}
        \begin{center}
                \includegraphics[width= \textwidth]{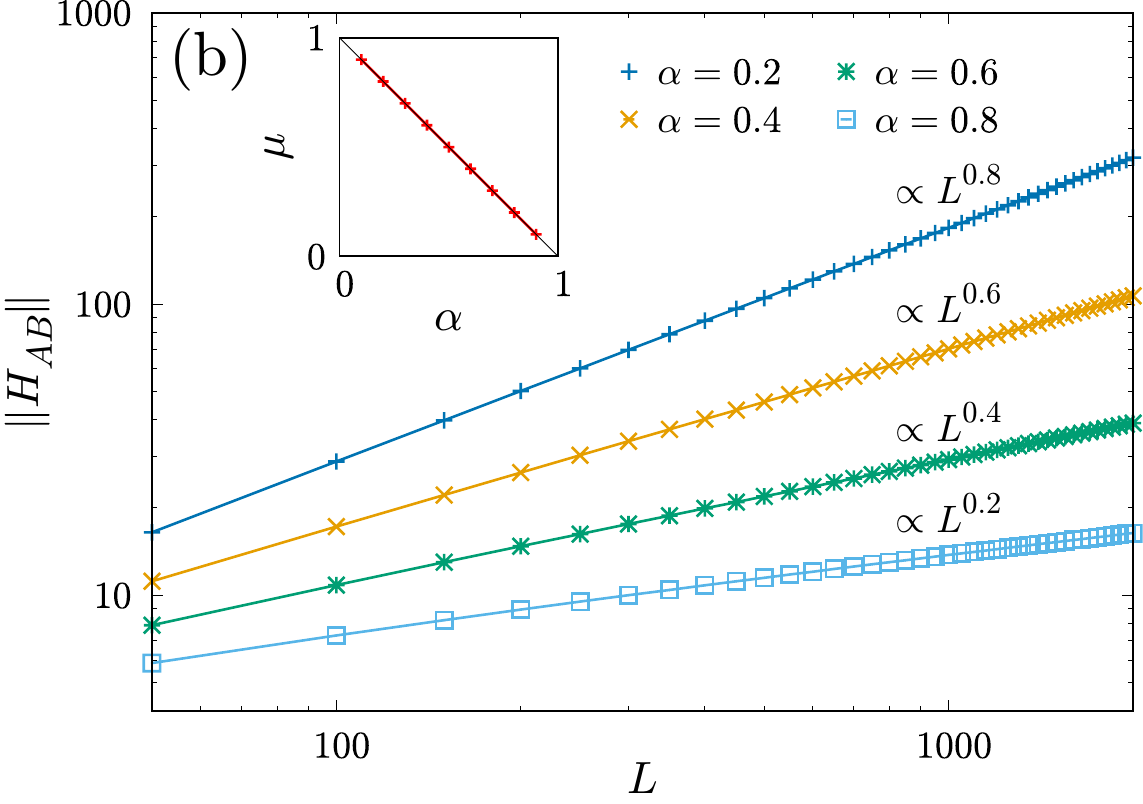}
        \end{center}
      \end{minipage}

      \begin{minipage}{0.34\hsize}
        \begin{center}
                \includegraphics[width= \textwidth]{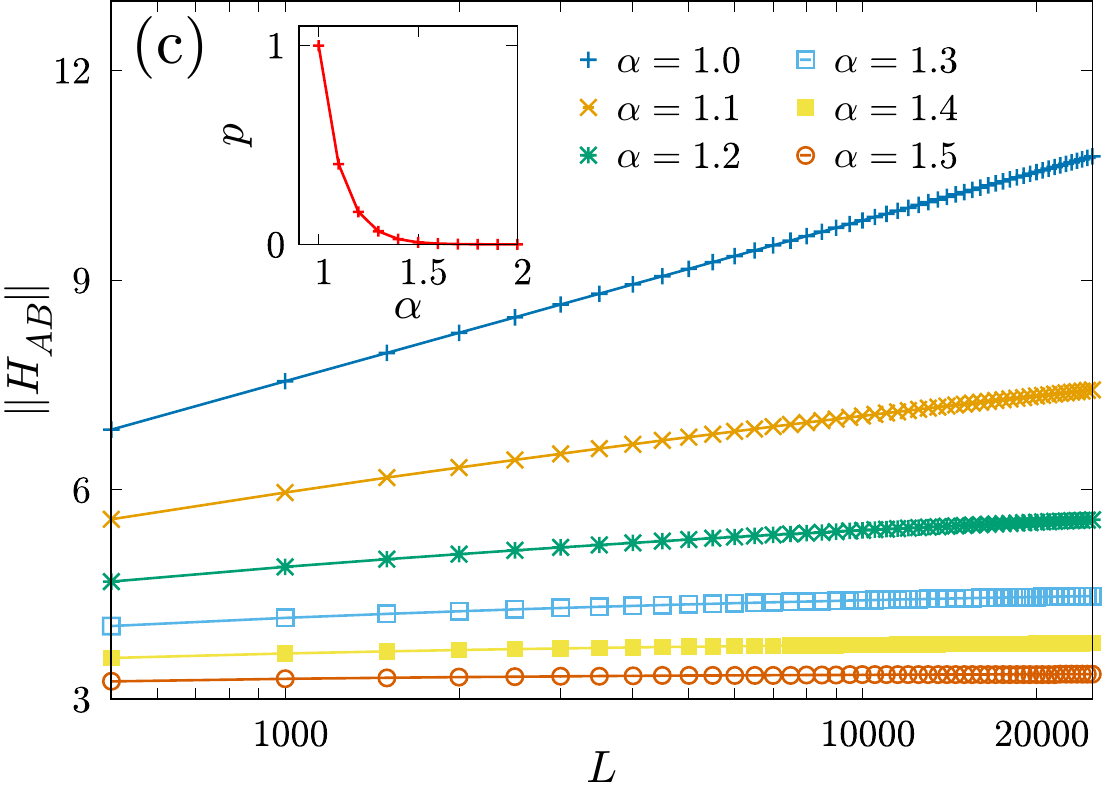}
        \end{center}
      \end{minipage}
    \end{tabular}
    \caption{\label{fig3:suppl} Operator norm of the boundary-interaction Hamiltonian $\| H_{AB} \|$ for $\ell=L/2$ and $V=0$. (a): The operator norm for various system-sizes as a function of $\alpha$. We show the data up to $L=20000$. We see that the norms are constants for $\alpha > 1.5$, which is guaranteed by the Lemma 1. (b): The operator norm as a function of the system size  for the regime $\alpha< 1$ in the log-log scale. The lines are fittings by the function $a L^{\mu} + b$. (c): The operator norm as a function of the system size for the regime $1<\alpha< 1.5$ in the semi-log scale. We show the data up to the system size $L=25000$. The lines are fittings by the function $p \log L + q$. From these data, the system-size dependence indicates that $\| H_{AB} \| \propto L^{1-\alpha}$ for $\alpha <1$,  $\propto \log L$ for $1<\alpha < 1.5 $ and constants for $\alpha >1.5$.}
      \end{figure}

\section{Supplementary numerical data for the interacting systems}
In this section, we consider the Hamiltonian
\begin{align}
  H &=
      \sum_{j=1}^L \sum_{r=1}^{L/2}{1\over r^{\alpha}} \left[ -c_{j+r}^{\dagger}c_j
      -c_{j}^{\dagger}c_{j+r} + V n_{j+r} n_j  \right] \, , \label{supplhamil}
\end{align}
where we set $V=1$. We consider the boundary-interaction Hamiltonian $H_{AB}$:
\begin{align}
  H_{AB}&:= \sum_{i\in A}\sum_{j \in B} h_{i,j} =
          \sum_{j=1}^{\ell} \left[\sum_{k=L/2+j}^L {1\over |L+j-k|^{\alpha}} + \sum_{k=\ell +1}^{j + L/2} {1\over |k-j|^{\alpha}}\right]   \left[ -c_j^{\dagger} c_k - c_k^{\dagger} c_j + V n_j n_k \right] \, .
\end{align}
We calculate the operator norm $\| H_{AB} \|$ using the density-matrix renormalization group technique \cite{White1992, White1993, Schollwock2011} implemented in the ITensor Library \cite{fishman2020itensor}. The results are shown in Fig.~\ref{supplsugifig2}, which shows the power-law dependence on the system size $L$.
We fitted the results $\| H_{AB}\| = a L^{\mu} + b$ at $L \geq 64$. The exponents $\mu$ are shown as a function of $\alpha$ in the inset, which indicates that
\begin{eqnarray}
  \| H_{AB}\| &\propto
                \left\{
                \begin{array}{ll}
                  L^{2 - \alpha } & ~~\alpha < 2  \, , \\
                  {\rm const.} &~~ \alpha > 2  \, .
                \end{array}
                \right. \label{suppl:habpower}
\end{eqnarray}
From this dependence, we anticipate that the MIP exists (at least) for $\alpha > 2$.

\begin{figure}[t]
\centering
{
  \includegraphics[clip, width=0.5\linewidth]{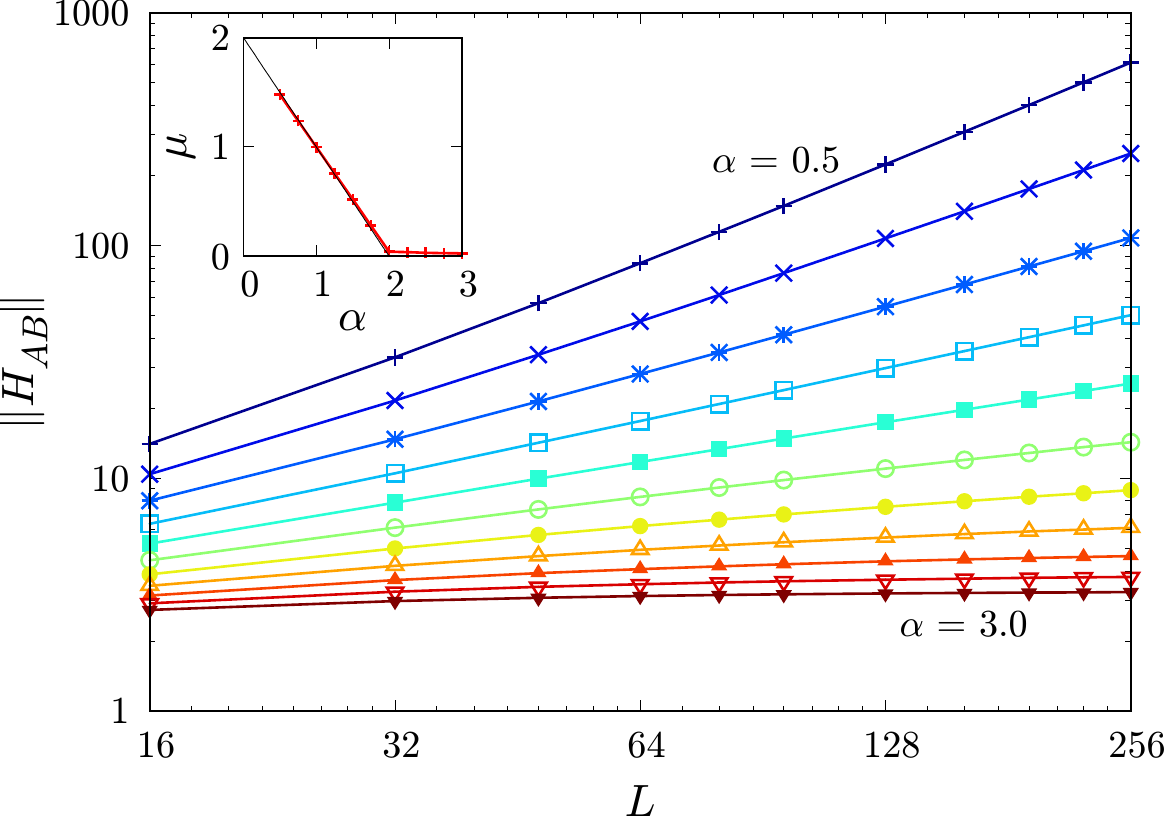}
  \caption{
    Size dependence of $\| H_{AB}\|$ at $\alpha = 0.5, 0.75, 1.0, \cdots, 3.0$. The numerical calculation shows power-law behavior (\ref{suppl:habpower}).
    \label{supplsugifig2}}
}

\end{figure}

We perform time-evolution calculations of the interacting systems using the one-site version of the time-dependent variational principle method \cite{Haegeman2011, Haegeman2013, Yang2020}. The truncation error is set to be smaller than $10^{-10}$, and after each measurement, we enlarge the bond dimension of the matrix-product state using the global-subspace expansion method \cite{Yang2020}. For each system size, we use the same number of trajectories as in the free fermion case \cite{ft0}.
In Fig.~\ref{supplsugifig1}, we show the entanglement entropy for $\alpha =3.0$ and $\alpha=0.5$, as typical $\ell$-dependence for different measurement amplitudes $\gamma$.  For the case of $\alpha=3.0$, sufficiently strong measurements suppress the entanglement growth, leading to the area law, that is, they become flat in small $\ell$ regimes. Conversely, for $\alpha=0.5$, the entanglement entropies never become flat before $L/2$, even for large measurement amplitudes. We remark that the entanglement entropy at $L/2$ is always flat, as seen in the Page curve for the random matrix.

\begin{figure}[t]
\centering
{
  \includegraphics[clip, width=0.5\linewidth]{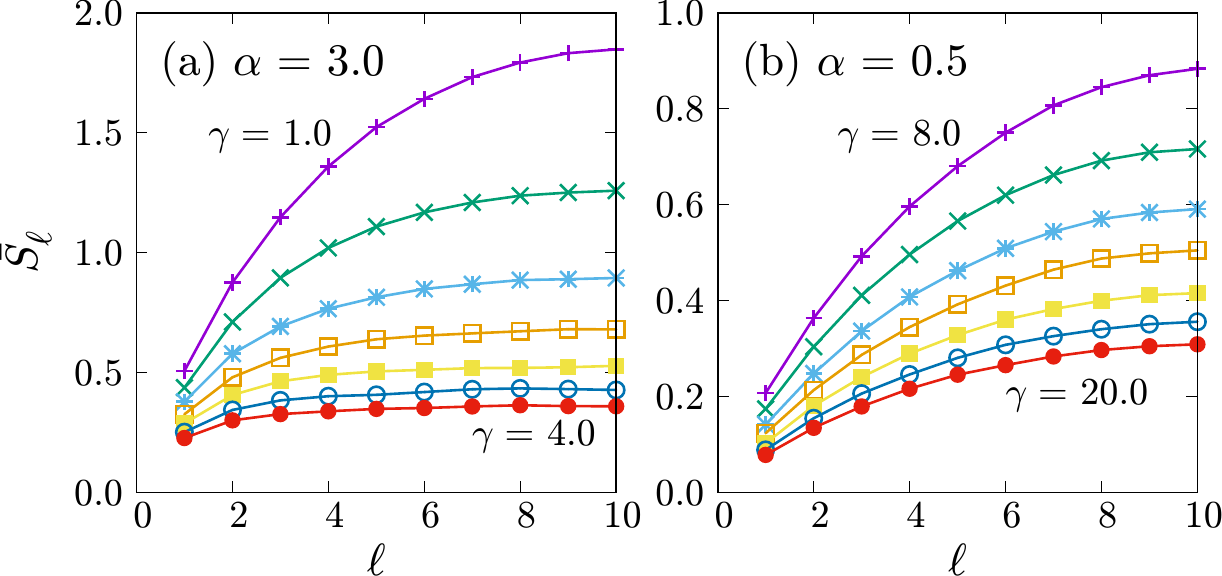}
  \caption{
    $\ell$-dependence of the entanglement entropy for various measurement amplitudes $\gamma$ at $L=20$. (a): $\alpha =3.0$ and $\gamma = 1.0, 1.5, 2.0, \cdots, 4.0$. (b): $\alpha=0.5$ and $\gamma = 8.0, 10.0, 12.0, \cdots, 20.0$.
    \label{supplsugifig1}
  }
}
\end{figure}

\end{widetext}

\end{document}